\documentclass[preprint]{iucr}              

\usepackage{amssymb}
\usepackage{epstopdf}
\usepackage{bmpsize}
\usepackage{pxfonts}
\usepackage{epstopdf}
\usepackage{bm}
\usepackage{graphicx}
\usepackage{color}

\journalcode{S}              

\begin{document}                  
	\thispagestyle{empty}
	\begin{Large}
		\textbf{DEUTSCHES ELEKTRONEN-SYNCHROTRON}
		
		\textbf{\large{in der HELMHOLTZ-GEMEINSCHAFT}\\}
	\end{Large}

	DESY 18-112 
	
	July 2018

	\begin{eqnarray}
	\nonumber
	\end{eqnarray}
	\begin{center}
		\begin{Large}
			\textbf{Effects of energy spread on Brightness and Coherence of undulator sources}
		\end{Large}
		\begin{eqnarray}
		\nonumber &&\cr 
		\end{eqnarray}
		
		\begin{large}
			Gianluca Geloni, Svitozar Serkez
		\end{large}
		\textsl{\\European XFEL, Schenefeld}
		\begin{eqnarray}
		\nonumber
		\end{eqnarray}
		\begin{large}
			Ruslan Khubbutdinov, Vitali Kocharyan and Evgeni Saldin
		\end{large}
		\textsl{\\Deutsches Elektronen-Synchrotron DESY, Hamburg}
		\begin{eqnarray}
		\nonumber
		\end{eqnarray}
		\begin{eqnarray}
		\nonumber
		\end{eqnarray}
		ISSN 0418-9833
		\begin{eqnarray}
		\nonumber
		\end{eqnarray}
		\begin{large}
			\textbf{NOTKESTRASSE 85 - 22607 HAMBURG}
		\end{large}
	\end{center}
	\clearpage
	\newpage

	
	
	\title{Effects of energy spread on Brightness and Coherence of undulator sources}
	
	
	\author[a]{Gianluca}{Geloni}
	\author[a]{Svitozar}{Serkez}
	\author[b]{Ruslan}{Khubbutdinov}
	\author[b]{Vitali}{Kocharyan}
	\author[b]{Evgeni}{Saldin}
	
	\aff[a]{European XFEL GmbH, Hamburg \country{Germany}}
	\aff[b]{Deutsches Elektronen Synchrotron, Hamburg \country{Germany}}
	
	
	
	
	
	
	
	
	
	\maketitle                        
	
	\begin{synopsis}
		The advent of "diffraction-limited" storage rings with ultra-low emittance poses a question on possible limitations to the (spectral) brightness and coherence due to the electron beam energy spread. We study this question using semi-analytical techniques.
	\end{synopsis}
	
	\begin{abstract}
		 The (spectral) brightness for partially transverse coherent sources as Synchrotron Radiation (SR) and Free-Electron Laser (FEL) sources can be defined as the maximum of the Wigner distribution. Then, the brightness includes information on both coherence and wavefront characteristics of the radiation field. For undulator sources, it is customary to approximate the single-electron electric field at resonance with a Gaussian beam, leading to great simplifications. Attempts to account for the modified spatial and angular profile of the undulator radiation in the presence of detuning due to energy spread currently build on the simplified brightness expression derived under the assumption of Gaussian beams. The influence of energy spread on undulator radiation properties is becoming important in view  of diffraction-limited rings with ultralow emittance coming on-line. Here we discuss the effects of energy spread on the brightness of undulator radiation at resonance, as well as relevant relations with coherence properties. 		
	\end{abstract}


\section{Introduction}

The concept of (spectral) brightness, which is used as a figure of merit for synchrotron radiation (SR) and FEL sources, is historically rooted in radiometry \cite{born99}. Radiometry treats radiation within the framework of geometrical optics and characterizes sources in terms of radiance, that is the maximum photon flux density in phase space, measured as a spectral photon flux per unit area per unit projection solid angle. Other quantities of interest can be derived by computing the marginals of the photon phase space distribution. A particularly attractive feature of the radiance is that for non-dissipative systems where the Liouville theorem holds, this quantity is an invariant along the beamline. Therefore, it is strictly related to the maximum spectral photon flux that can be obtained at the sample position, assuming an ideal optical system. 

Starting from the pioneering works \cite{coiss86,kim86,kim87,kim87b},  a lot of literature is available, which deals with  the generalization of the concept of radiance to the case of partially transverse coherent sources as Synchrotron Radiation (SR) and Free-Electron Laser (FEL) sources \cite{coiss86,kim86,kim87,kim87b,hulbert92,howells94,bahrdt97,hulbert98,hulbert98b,bosch99,attwood99,bosch00,ciocci00,duke00,xraydata01,wiedemann02,onuki03,hofmann04,clarke04,talman06,williams06,tanaka09,bazarov12,huang13,tanaka14,geloni15,varta18}. This generalization process required changing the working framework from pure geometrical optics to  wave optics, backed up by statistical optics. This led to the substitution of the phase space of optical rays in geometrical optics with a Wigner distribution that, as is well-known, is a quasi-probability distribution, not everywhere positive definite. 

As underlined in \cite{bazarov12}, this generalization process naturally includes a strong analogy with quantum mechanics in position representation, where wave functions are analogous to spatial field distributions and the classical concept of phase space is substituted by a Wigner distribution, which can assume negative values related to the ability of wave functions (and electric fields) to interfere. In quantum mechanics (or in wave optics) one often deals with random processes so that it becomes necessary to describe the state of the system (or the electric field) in terms of a density matrix, which assumes the form of a correlation function in position representation. In the case for SR and FEL sources the underlying, fundamental  stochastic process  is the shot-noise in the electron beam. As is well-known, in statistical optics the spatial field correlation function at a given frequency takes the name of cross-spectral density. The overall degree of transverse coherence is just analogous to the trace of the squared of the density matrix representing the statistical operator for a quantum mixture, and can therefore be expressed in terms of integrals of the cross-spectral density.  It is interesting to remark here that the trace of a matrix is invariant with respect to a basis transformation. This fact is well-known in statistical quantum mechanics, where a mixed state can be thought as a mixture of pure states that diagonalize the statistical operator with weights given by its eigenvalues. The same fact is similarly well-known in statistical optics, where the coherent mode decomposition theorem allows to write a cross-spectral density as a sum of uniquely defined statistically independent contributions, obviously leaving the overall degree of coherence unvaried.

The relation between cross-spectral density (or density matrix) and Wigner distribution is a simple Fourier transformation. In other words, they carry the same identical amount of information. The brightness can be seen as a figure of merit that is extracted from the Wigner distribution.  There are several recipes for doing so. One defines the brightness in terms of integrals of the Wigner function and of its square. Another defines it as the Wigner function on axis. See, for example, \cite{bazarov12} for a review.

However, as noted in \cite{geloni15}, there is a correspondence principle between wave and geometrical optics, exactly as there is a correspondence principle between quantum and classical mechanics. In particular, there is a special class of sources, called quasi-homogeneous sources, for which the Wigner distribution function factorizes as

\begin{eqnarray}
W(\bm{r}, \bm{\theta}) = I(\bm{r}) \mathcal{I}(\bm{\theta})~,
\label{Wig2}
\end{eqnarray}
where $ I(\bm{r}) $ and  $\mathcal{I}(\bm{\theta})$  can be respectively identified with the source intensity distribution and with the angular distribution of radiation intensity.  Then, $W$ is the product of two positive quantities, and, being positive-definite everywhere, can be identified with a phase space. In this limit, the brightness must strictly correspond to the radiance, and is the maximum of the Wigner distribution function. It is therefore natural to define the brightness for any source as the maximum of the Wigner distribution.  

With this last definition, the brightness includes information on both coherence and wavefront characteristics of the electric field, in contrast to the case where it is defined in terms of integrals of the Wigner function, and only information on the degree of coherence is present.

The previous discussion is meant to be a quick summary of the strict relations between coherence properties and brightness, which are important to keep in mind when discussing about radiation properties from FELs and storage ring sources, and is becoming more important for storage-ring based sources, in view of the coming on-line of many state-of-the-art diffraction-limited rings.

For the case of storage rings, one can approximate the transverse electron phase space with a Gaussian function. Moreover, for undulator sources, it is customary to approximate the single-electron electric field at resonance with a Gaussian beam. In contrast to the real undulator field, Gaussian functions are separable, and a simplified expression for the brightness results in this case \cite{kim86}

\begin{eqnarray}
B= \frac{F}{4 \pi^2 \Sigma_x \Sigma_y \Sigma_{x'} \Sigma_{y'}}~,
\label{Busual}
\end{eqnarray}
where $F$ indicates the total flux per unit spectral bandwidth, while $\Sigma_{x,y}$, $\Sigma_{x',y'}$ are effective source size and divergences, calculated by summing in quadrature the sizes and divergences of the electron beam and of the single-electron radiation. 

Eq. (\ref{Busual}), derived under the Gaussian beam approximation,  does not include detuning or energy spread effects on the radiation beam. However, for diffraction-limited rings, studying the influence of energy spread of undulator radiation properties is becoming more and more important, because of the ultra-low electron emittance.

In \cite{tanaka09} an attempt is reported where the authors account for the modified spatial and angular profile of the undulator radiation in the presence of detuning. However, the approximate formula for the brigthness that they obtain still builds on Eq. (\ref{Busual}), that is based on the Gaussian beam approximation in the first place.  

It is therefore interesting to study energy spread effects on the brightness of undulator radiation by avoiding to rely on the Gaussian beam approximation from the very beginning, and defining the brightness as the maximum of the Wigner distribution. Moreover, given the strict relation between coherence and brightness, highlighted above, one should complement a study on the effects on the brightness with a study on the effects on coherence.

Here we will discuss the effects of energy spread on both  brightness and coherence of undulator radiation at resonance. We will first introduce basic quantities and notations. Then, using a simple model we will show a very counter intuitive fact. In the limit for a vanishing small emittance the brightness from an undulator is not influenced by the electron beam energy spread, in the case of a symmetrical distribution around the nominal energy. Further on, with the help of semi-analytical calculations, we will discuss the impact of energy spread on coherence properties of undulator radiation. We will illustrate our results with examples compatible to modern diffraction-limited sources, discussing similarities and differences with respect to the approach in \cite{tanaka09}.

\section{Basic quantities and notations \label{sec:2}}
	
We follow notations similar to \cite{geloni08} and \cite{geloni15}. We write the fundamental wavelength of a planar undulator with $N_u \gg 1$ periods as $\lambda_1 = \lambda_u (1+K^2/2)/(2 \gamma_1^2)$, where $\lambda_u$ is the undulator period, $L_u = N_u \lambda_u$, $k_u = 2\pi/\lambda_u$ and $K$ the maximum undulator parameter. Likewise, the fundamental frequency is $\omega_1 = 2 \pi c /\lambda_1$.  $\bm{\bar{E}}(\omega)$ denotes the Fourier transform of the electric field, and we define with $\bm{\widetilde{E}}(\omega) = \bm{\bar{E}}(\omega) \exp(-i\omega z/c)$ the slowly varying envelope of the field in the frequency domain, which we will refer to simply as "the field". 

Consider an electron that enters the undulator at a small angle $\bm{\eta}$ and offset $\bm{l}$ with energy fixed by $\gamma$ that can be different from the nominal value $\gamma_1$. The far field angular distribution seen at a distance $z \gg L_u$ from the middle of the undulator and at frequency $\omega$ such that $|\omega - \omega_1|/\omega_1 \ll 1$ (where the resonance approximation applies) depends on the parameters $z, \gamma, \bm{\eta}$ and $\bm{l}$, and is given by\footnote{Note the minus sign in the first term under the sinc function. If, for example, we fix $\omega = \omega_1$ but our electron has $\gamma > \gamma_1$, then the resonance frequency for that electron is higher than $\omega_1$, effectively corresponding to a negative detuning $-2 \pi N_u (\gamma-\gamma_1)/\gamma_1$.} 

\begin{eqnarray}
\widetilde{{E}}\left(\bm{\theta}\right)= -\frac{K \omega e L_u A_{JJ}} {2 c^2 z \gamma}
\exp\left[i\frac{\omega}{c}\left( \frac{z \theta^2}{2}-
\bm{\theta}\cdot\bm{l} \right)\right]
\mathrm{sinc}\left[-\frac{2 \pi N_u (\gamma-\gamma_1)}{\gamma_1} + \frac{\omega L_u
	\left|\bm{\theta}-\bm{\eta}\right|^2}{4 c}\right] ~.
\label{undurad4bisgg}
\end{eqnarray}
Here $A_{JJ} = J_0[K^2/(4+2K^2)]-J_1[K^2/(4+2K^2)]$ is the coupling strength for the first harmonic under the resonance approximation. Our considerations can be easily extended to odd harmonics. For even harmonics one should consider, instead, a different position of the maximum of the Wigner function. Note that under the resonance approximation the field is linearly polarized, hence $\widetilde{{E}}$ is a scalar quantity. An expression for the field at the virtual position $z=0$ i.e. in the middle of the undulator and for any position after the undulator exit at perfect resonance can be found in  Eq. (34) and Eq. (35) of reference \cite{geloni07}. However, to the authors' knowledge there is no analytical expression for the field at $z=0$ at finite detuning, which should be calculated propagating Eq. (\ref{undurad4bisgg}).

Following the references above we use normalized units defined as

\begin{eqnarray}
&& \bm{\hat{\eta}} = \frac{\bm{\eta}}{\sqrt{\lambdabar/L_u}} ~,~~~     \bm{\hat{\theta}} = \frac{\bm{\theta}}{\sqrt{\lambdabar/L_u}} ~,\cr
&& \bm{\hat{r}} = \frac{\bm{r}}{\sqrt{\lambdabar L_u}} ~,~~~~~\bm{\hat{l}} = \frac{\bm{l}}{\sqrt{\lambdabar L_u}} ~,\cr
&& \hat{\phi} = \frac{c t}{\lambdabar} ~,~~~~~~~~~~
 \hat{\xi}_E = -4 \pi N_u \frac{\gamma-\gamma_1}{\gamma_1} ~
\label{normq}
\end{eqnarray}
so that it is natural to define

\begin{eqnarray}
&& N_{x,y} = \frac{\sigma_{x,y}^2}{\lambdabar L_u}  ~,~~~
D_{x,y} = \frac{\sigma_{x',y'}^2}{\lambdabar/L_u}~, \cr&&
\Delta_\phi = \left(\frac{c\sigma_t}{\lambdabar}\right)^2~,~~
\Delta_E = \left(4 \pi N_u \sigma_{\Delta\gamma/\gamma}\right)^2~.
\label{ND2}
\end{eqnarray}
Roughly speaking, this amounts to normalizing angles to the diffraction angle  of single-electron emission, sizes to the diffraction size, fractional energy deviation to the undulator resonant bandwidth, and times to inverse radiation frequency. Moreover, here $\sigma_{x,y,t,\Delta\gamma/\gamma}$ are the rms of the electron beam dimensions in phase space, and we assume for simplicity that at $z=0$, i.e. in the middle of the undulator, the electron beam phase space can be factorized as

\begin{eqnarray}
\hat{f}_{6D} = f_{\eta_x}(\hat{\eta}_x) f_{\eta_y}(\hat{\eta}_y)
f_{l_x}(\hat{l}_y) f_{l_x}(\hat{l}_y) f_{\phi}(\hat{\phi}) f_{\xi_E}(\hat{\xi}_E)~,
\label{fphase}
\end{eqnarray}
with

\begin{eqnarray}
&& f_{\eta_x}(\hat{\eta}_x) = \frac{1}{\sqrt{2\pi D_x}}
\exp{\left(-\frac{\hat{\eta}_x^2}{2 D_x}\right)}~,~~~~
f_{\eta_y}(\hat{\eta}_y)  = \frac{1}{\sqrt{2\pi D_y}}
\exp{\left(-\frac{\hat{\eta}_y^2}{2 D_y}\right)}~,\cr &&
f_{l_x}(~\hat{l}_x) =\frac{1}{\sqrt{2\pi N_x} }
\exp{\left(-\frac{\hat{l}_x^2}{2 N_x}\right)}~,~~~~~
f_{l_y}(~\hat{l}_y)=\frac{1}{\sqrt{2\pi N_y} }
\exp{\left(-\frac{\hat{l}_y^2}{2 N_y}\right)}~,\cr &&
f_{\phi}(\hat{\phi})  = \frac{1}{\sqrt{2\pi \Delta_\phi}}
\exp{\left(-\frac{\hat{\phi}^2}{2 \Delta_\phi}\right)}~,~~~~
f_{\xi_E}(\hat{\xi}_E)  = \frac{1}{\sqrt{2\pi \Delta_E}} \exp{\left(-\frac{\hat{\xi}^2}{2 \Delta_E}\right)}~, \label{distr}
\end{eqnarray}
where we defined the various Gaussian distributions in terms of the variances $N_{x,y}$, $D_{x,y}$, $\Delta_E$ and $\Delta_\phi$ and we introduced $f_\phi(\hat{\phi})$ and $\Delta_\phi$ only for completeness, because in this paper we deal with spontaneous radiation and therefore these quantities are not used. The far-zone field in normalized units can be written as

\begin{eqnarray}
\hat{E}(\bm{\hat{\theta}}) = \frac{1}{\hat{z}} \exp\left[i \frac{\hat{\theta}^2\hat{z}}{2} - i \bm{\hat{\theta}}\cdot \bm{\hat{l}}\right] \mathrm{sinc}\left(\frac{\hat{\xi}_E}{2} + \frac{|\bm{\hat{\theta}}-\bm{\hat{\eta}}|^2}{4}\right) ~,\label{fffiledn}
\end{eqnarray}
where $\hat{z} = z/L_u$. As discussed above, one may calculate the analogous field at the virtual source as

\begin{eqnarray}
\hat{E}(\bm{\hat{r}}) =-i \exp\left[i \bm{\hat{\eta}} \cdot \left(\bm{\hat{r}}-\bm{\hat{l}}\right)\right] \int_0^\infty d \hat{\theta}~\hat{\theta} J_0\left(\hat{\theta}\left|\bm{\hat{r}}-\bm{\hat{l}}\right|\right)  \mathrm{sinc}\left(\frac{\hat{\xi}_E}{2} + \frac{\hat{\theta}^2}{4}\right)\label{nnfiledn}
\end{eqnarray}

\begin{figure}
	\centering
	\includegraphics[width=0.75\textwidth]{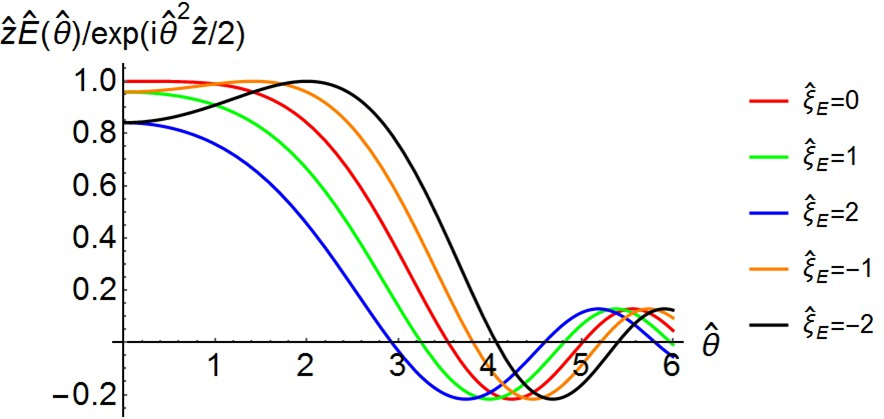}
	\\~\\
	\includegraphics[width=0.75\textwidth]{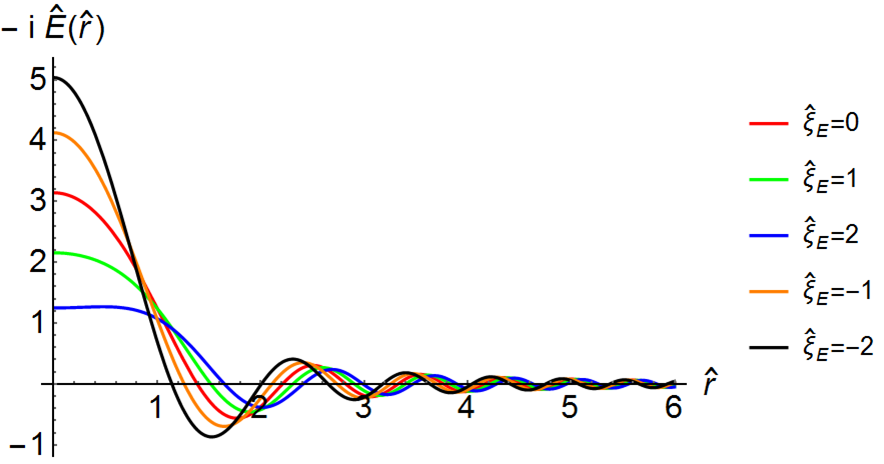}
	\caption{Top panel: the function $\hat{z}\hat{E}(\hat{\theta})/\exp(i \hat{\theta}^2\hat{z}/2)$ is plot for different values of $\hat{\xi}_E$. Bottom panel: the field at the virtual source located in the middle of the undulator, $-i \hat{E}(\hat{r})$ is plot for different values of $\hat{\xi}_E$. Both functions are axis-symmetric, i.e. a 3D picture can be obtained by a rotation around the vertical axis. Here $\bm{\hat{l}}=\bm{0}$ and $\bm{\hat{\eta}}=\bm{0}$.}
	\label{uno}
\end{figure}

\begin{figure}
	\centering
	\includegraphics[width=0.75\textwidth]{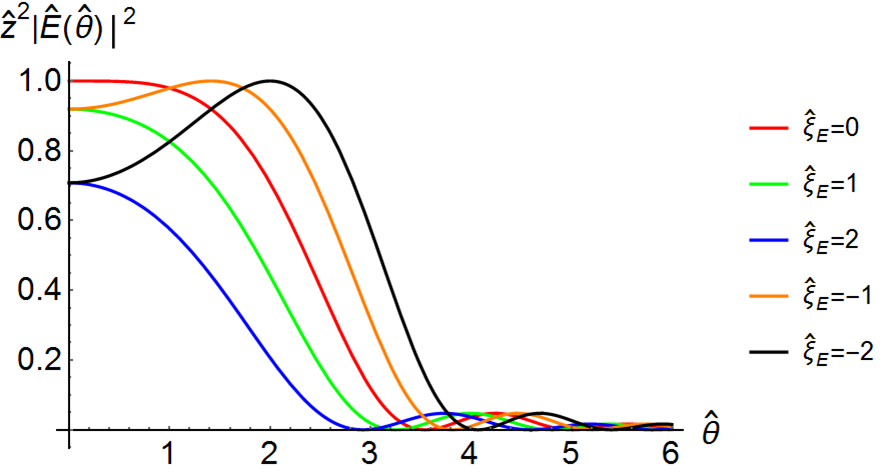}
	\\~\\
	\includegraphics[width=0.75\textwidth]{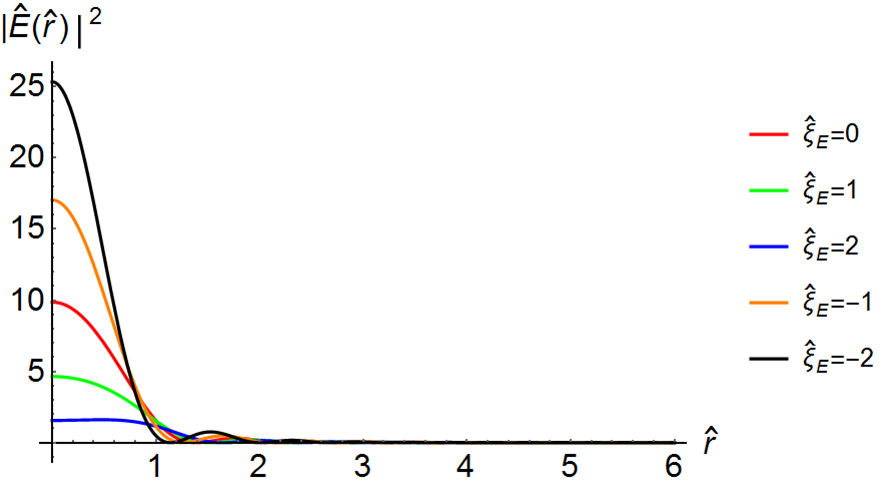}
	\caption{Top panel: the function $\hat{z}^2  |\hat{E}(\hat{\theta})|^2$ is plot for different values of $\hat{\xi}_E$. Bottom panel: the function  $|\hat{E}(\hat{r})|^2$ is plot for different values of $\hat{\xi}_E$. Both functions are axis-symmetric, i.e. a 3D picture can be obtained by a rotation around the vertical axis. Here $\bm{\hat{l}}=\bm{0}$ and $\bm{\hat{\eta}}=\bm{0}$.}
	\label{unob}
\end{figure}
In the top panel of Fig. \ref{uno} we plot $\hat{z}\hat{E}(\hat{\theta})/\exp(i \hat{\theta}^2\hat{z}/2)$, i.e. the well-known far-field profile for $\bm{\hat{l}}=\bm{0}$ and $\bm{\hat{\eta}}=\bm{0}$ (which is azimuthal-symmetric) as a function of $\hat{\theta}$ for different values of the detuning $\hat{\xi}_E$, while on the bottom panel we plot the field at the virtual source (which has a plane wavefront) for the same choices of the detuning parameter.
For comparison, in the top and bottom panels of Fig. \ref{unob} we plot, respectively, $\hat{z}^2  |\hat{E}(\hat{\theta})|^2$ and $|\hat{E}(\hat{r})|^2$ that are the corresponding intensity distributions.

It can be shown that for negative values of $\hat{\xi}_E$ the maximum intensity at the source increases and tends to "saturate" for large negative values, while it remains constant in the far zone. We will discuss the consequences of this fact later on. 

Also, even at $\hat{\xi}_E=0$, the intensity distribution at the virtual source and in the far zone are not Gaussian. Therefore, any Gaussian approximation relies on a fitting procedure. In this regard it is important to remark that the intensity distribution in the far zone and at the virtual source are related by the laws of field propagation in free-space, basically a Fourier transformation. One may fit the  intensity at the virtual source with a Gaussian, but in that case the real intensity in the far-zone does not match the propagated Gaussian beam. One may fit the intensity in the far zone with the Gaussian, but in that case the intensity at the virtual source does not match the back-propagated Gaussian beam. In other words, there is some freedom when it comes to apply the Gaussian approximation. Many different choices can be found in literature, see for example \cite{kim86,kim87,lindberg15}. One of the possible choices \cite{kim87,tanaka09} is to fix, for the single-electron intensity distribution, $\sigma_r =  \sqrt{2\lambda L_u}/(4 \pi)$ and  $\sigma_{r'} = \sqrt{\lambda/(2 L_u)}$, corresponding to the photon emittance (strictly speaking we cannot define a photon emittance, except in those cases when the Wigner distribution is positive definite, and the Gaussian approximation is one of those cases) $\epsilon_r = \sigma_r \sigma_{r'}=\lambda/(4\pi)$. In our normalized units, they amount to $\hat{\sigma}_r = 1/(2\sqrt{\pi})$ and $\hat{\sigma}_{r'} = \sqrt{\pi}$. The corresponding FWHM values are obtained multiplying by $2\sqrt{\ln 2}\simeq 2.35$ and read $\delta\hat{r}_{Gauss} = 0.664$ and $\delta\hat{r'}_{Gauss}=4.17$, to be compared with the corresponding FWHM values for the actual intensities at $\hat{\xi}_E=0$, which are found to be $\delta\hat{r}_{real} =1.36$ and $\delta\hat{r'}_{real}=4.72$.

Having discussed the single-electron field and intensity profiles, we now introduce the  cross-spectral density in normalized units: 

\begin{eqnarray}
\hat{G}(\bm{\hat{\theta}}, \Delta \bm{\theta}) = \left\langle\hat{E}(\bm{\hat{\theta}}+\Delta \bm{\hat{\theta}}/2) \hat{E}^*(\bm{\hat{\theta}}-\Delta \bm{\hat{\theta}}/2)\right\rangle \label{big}
\end{eqnarray}
where the brackets $\langle..\rangle$ indicate averaging over an ensemble realizations, $\bm{\hat{\theta}}$ is the vector position at which a two-pinholes system is introduced to probe coherence, and $\Delta \bm{\hat{\theta}}$ is the vector describing the separation between the two pinholes, see Eq. (\ref{normq}). Clearly,  $\bm{\hat{\theta}}$ and $\Delta \bm{\hat{\theta}}$ may have different directions. We remind that the spectral degree of coherence is defined as

\begin{eqnarray}
g(\bm{\hat{\theta}}, \Delta \bm{\hat{\theta}}) = \frac{\hat{G}(\bm{\hat{\theta}},\Delta \bm{\hat{\theta}})}{\left[\hat{G}(\bm{\hat{\theta}}+\Delta \bm{\hat{\theta}}/2)\hat{G}(\bm{\hat{\theta}}-\Delta \bm{\hat{\theta}}/2)\right]^{1/2}}~,
\end{eqnarray}
and the fringe visibility of an interference experiment is given by

\begin{eqnarray}
V =  \frac{2 |\hat{G}(\bm{\hat{\theta}}, \Delta \bm{\hat{\theta}})|}{\hat{G}(\bm{\hat{\theta}}+ \Delta \bm{\hat{\theta}}/2,\bm{0})+\hat{G}(\bm{\hat{\theta}}- \Delta \bm{\hat{\theta}}/2,\bm{0})}~.
\label{visi}
\end{eqnarray}
Finally, the Wigner distribution in normalized units is

\begin{eqnarray}
&& {\hat{W}(\bm{\hat{r}},\bm{\hat{\theta}})} =  \int d^2( \Delta \hat{{\theta}} ) \exp\left(i  {\bm{\hat{r}}} \cdot \Delta \bm{{\hat{\theta}}}\right)   \hat{G}(\bm{\hat{\theta}}, \Delta \bm{\hat{\theta}})~.
\label{Wigfarfin}
\end{eqnarray}
Following the same formalism as in \cite{geloni15}, the corresponding result in dimensional units is found  to be linked to Eq. (\ref{Wigfarfin}) by the constant 

\begin{eqnarray}
\mathcal{C} = \frac{z^2 I K^2 \omega^3 \alpha  A_{JJ}^2 }{64 \pi^4 e c^3 \gamma^2    L_u }
\end{eqnarray} 
~ with $\alpha = e^2/(\hbar c)$ the fine structure constant. This result follows from the correspondence principle for quasi-homogeneous sources discussed in the introduction, for which Eq.(\ref{Wig2}) is valid. The brightness, defined by us as the maximum of the Wigner distribution, is therefore given by

\begin{eqnarray}
B=\mathcal{C} \max(\hat{W})~.
\end{eqnarray}
Here we underline the fact that, while this is often the case, in the most general case the maximum of the Wigner function may not be on-axis, i.e. may not be at ${\hat{r}}=0$ and ${\hat{\theta}}=0$. Choosing the maximum of the Wigner function for defining the brightness assures that the correspondence principle discussed in the introduction is consistently applied.

Substitution of Eq. (\ref{fffiledn}) into Eq. (\ref{big}) gives the following explicit expression for the cross-spectral density in the case of undulator radiation around the fundamental harmonic (or, with simple changes, for odd harmonics)

\begin{eqnarray}
&&\hat{G}(\bm{\hat{\theta}}, \Delta \bm{\hat{\theta}}) = \frac{1}{(2\pi)^{3/2}\sqrt{D_x D_y\Delta_E} \hat{z}^2}\exp\left(-i \hat{z}\bm{\hat{\theta}} \cdot{\Delta \bm{\hat{\theta}}}\right) \exp\left(-\frac{N_x \Delta \hat{\theta}_x^2}{2}\right)\exp\left(-\frac{N_y \Delta \hat{\theta}_y^2}{2}\right) \cr && \times\int_{-\infty}^{\infty} d \hat{\eta}_x \int_{-\infty}^{\infty} d \hat{\eta}_y \int_{-\infty}^{\infty} d \hat{\xi}_E \exp\left(-\frac{\hat{\eta}_x^2}{2D_x}\right)\exp\left(-\frac{\hat{\eta}_y^2}{2D_y}\right)\cr && \times\exp\left(-\frac{\hat{\xi}_E^2}{2\Delta_E}\right) \mathrm{sinc}\left(\frac{\hat{\xi}_E}{2} + \frac{|\bm{\hat{\theta}}-\bm{\hat{\eta}}+\Delta \bm{\hat{\theta}}/2|^2}{4}\right) \mathrm{sinc}\left(\frac{\hat{\xi}_E}{2} + \frac{|\bm{\hat{\theta}}-\bm{\hat{\eta}}-\Delta \bm{\hat{\theta}}/2|^2}{4}\right) ~. \label{big0}
\end{eqnarray}

Note that the single-electron spectral-angular intensity distribution has, in our case, its maximum at resonance on axis. Then, for a Gaussian distribution of energy spread, divergence and size of the electron beam, the maximum of the Wigner distribution must be at $\hat{r}=0$ and $\hat{\theta}=0$, and therefore

\begin{eqnarray}
{B} = \mathcal{C} \cdot \hat{W}(\bm{0},\bm{0})  =  \mathcal{C} \int d^2( \Delta \hat{{\theta}} )    \hat{G}(\bm{0}, \Delta \bm{\hat{\theta}})~,
\label{Wigfarfin00}
\end{eqnarray}
the integral extending over all the plane spanned by the vector $\Delta \bm{\hat{\theta}}$.

\section{Effects of energy spread on the brightness \label{sec:3}}

Let us first consider the simplest case of a beam with vanishing emittance. Eq. (\ref{big0}) simplifies accordingly, and substitution into Eq. (\ref{Wigfarfin00}) gives the following expression for the brightness\footnote{Mathematically speaking, here we take limit for $N_{x,y} \longrightarrow 0$ and $D_{x,y} \longrightarrow 0$ .}

\begin{eqnarray}
&& {B} =  \frac{\sqrt{2 \pi} \mathcal{C}}{ \sqrt{\Delta_E} \hat{z}^2} \int_0^\infty d( \Delta \hat{{\theta}} )  \Delta \hat{\theta} \int_{-\infty}^{\infty} d \hat{\xi}_E~  \exp\left(-\frac{\hat{\xi}_E^2}{2\Delta_E}\right) \mathrm{sinc}^2\left(\frac{\hat{\xi}_E}{2} + \frac{(\Delta \hat{\theta}/2)^2}{4}\right)~,
\label{Wigfarfinboh}
\end{eqnarray} 
where we used the fact that in the limit for zero emittance $\hat{G}(\bm{0}, \Delta \bm{\hat{\theta}})$ is azimuthal symmetric. Now we note that 

\begin{eqnarray}
&& {B} =  \frac{\sqrt{2 \pi} \mathcal{C}}{ \sqrt{\Delta_E} \hat{z}^2} \int_{-\infty}^{\infty} d \hat{\xi}_E~  \exp\left(-\frac{\hat{\xi}_E^2}{2\Delta_E}\right) F(\hat{\xi}_E)~,
\label{Wigfarfinboh2}
\end{eqnarray} 
where

\begin{eqnarray}
F(\hat{\xi}_E) = \int_0^\infty d( \Delta \hat{{\theta}} )  \Delta \hat{\theta}  \mathrm{sinc}^2\left(\frac{\hat{\xi}_E}{2} + \frac{(\Delta \hat{\theta}/2)^2}{4}\right)  = \frac{4}{\hat{\xi}_E} \left[2+\pi \hat{\xi}_E - 2 \cos(\hat{\xi}_E) -2\hat{\xi}_E \mathrm{Si}(\hat{\xi}_E)\right]
\label{calcul}
\end{eqnarray} 
with $\mathrm{Si}(\hat{\xi}_E)= \int_0^{\hat{\xi}_E} dt ~\mathrm{sinc}(t)$ is the sine integral function.

By definition, the function $F(\hat{\xi}_E)$ is proportional to the angle-integrated spectral flux from a single electron, and therefore the brightness is proportional to the single-electron angle-integrated spectral flux, averaged over the energy spread distribution. 

Moreover, the function $F(\hat{\xi}_E)$ has the property that $F(\hat{\xi}_E)+F(-\hat{\xi}_E) =8 \pi$ independently of the value of the real number $\hat{\xi}_E$.  We conclude that for zero emittance and symmetric energy spread distribution we cannot have any effect of the energy spread on the brightness that can in fact be written as

\begin{eqnarray}
{B} = \frac{ I K^2 \omega^3 \alpha  A_{JJ}^2 L_u}{8 \pi^2 e c^3 \gamma^2    } ~. 
\label{Wigfarfinbohfin}
\end{eqnarray} 

\begin{figure}
	\centering
	\includegraphics[width=0.40\textwidth]{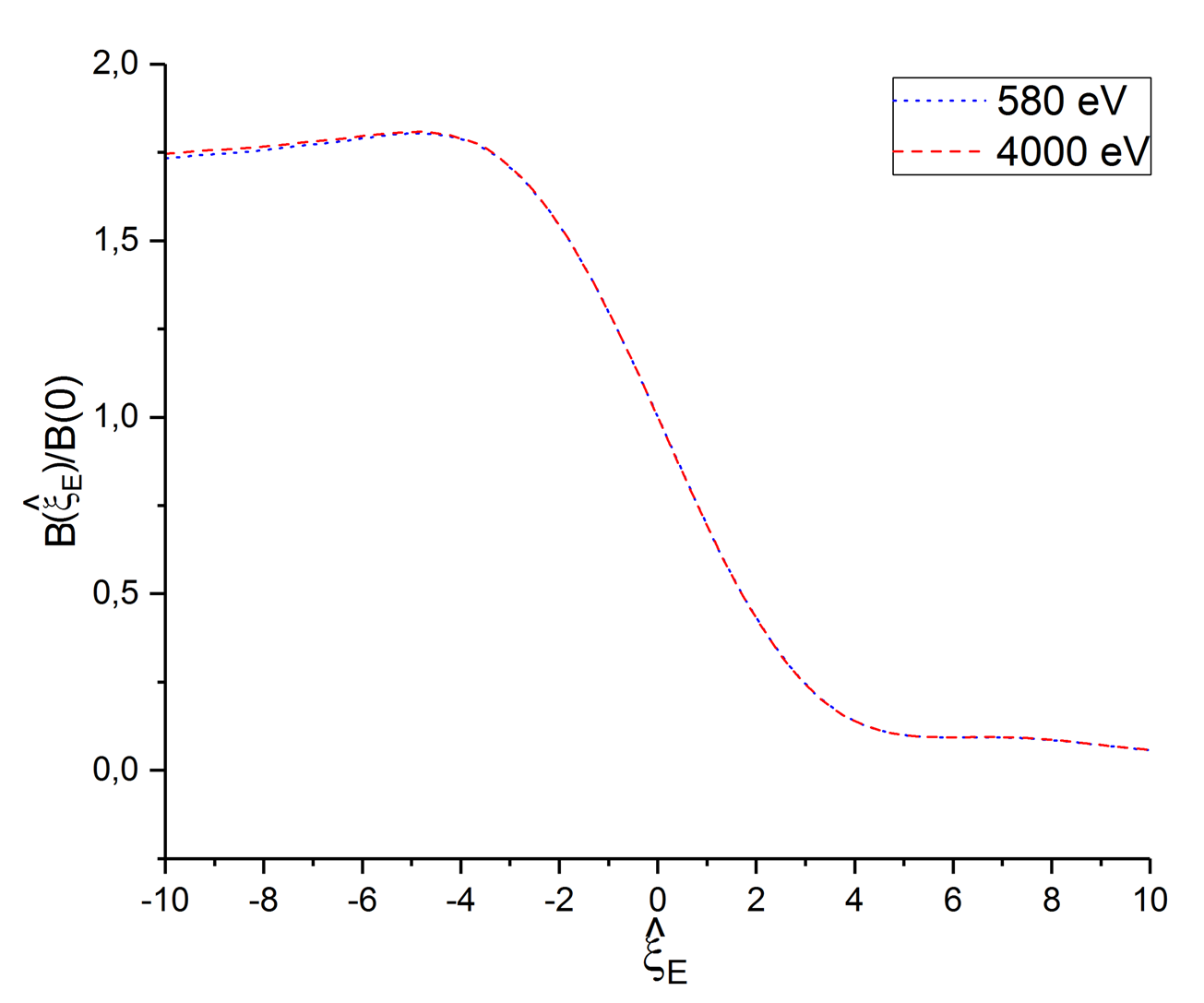}
	\includegraphics[width=0.5\textwidth]{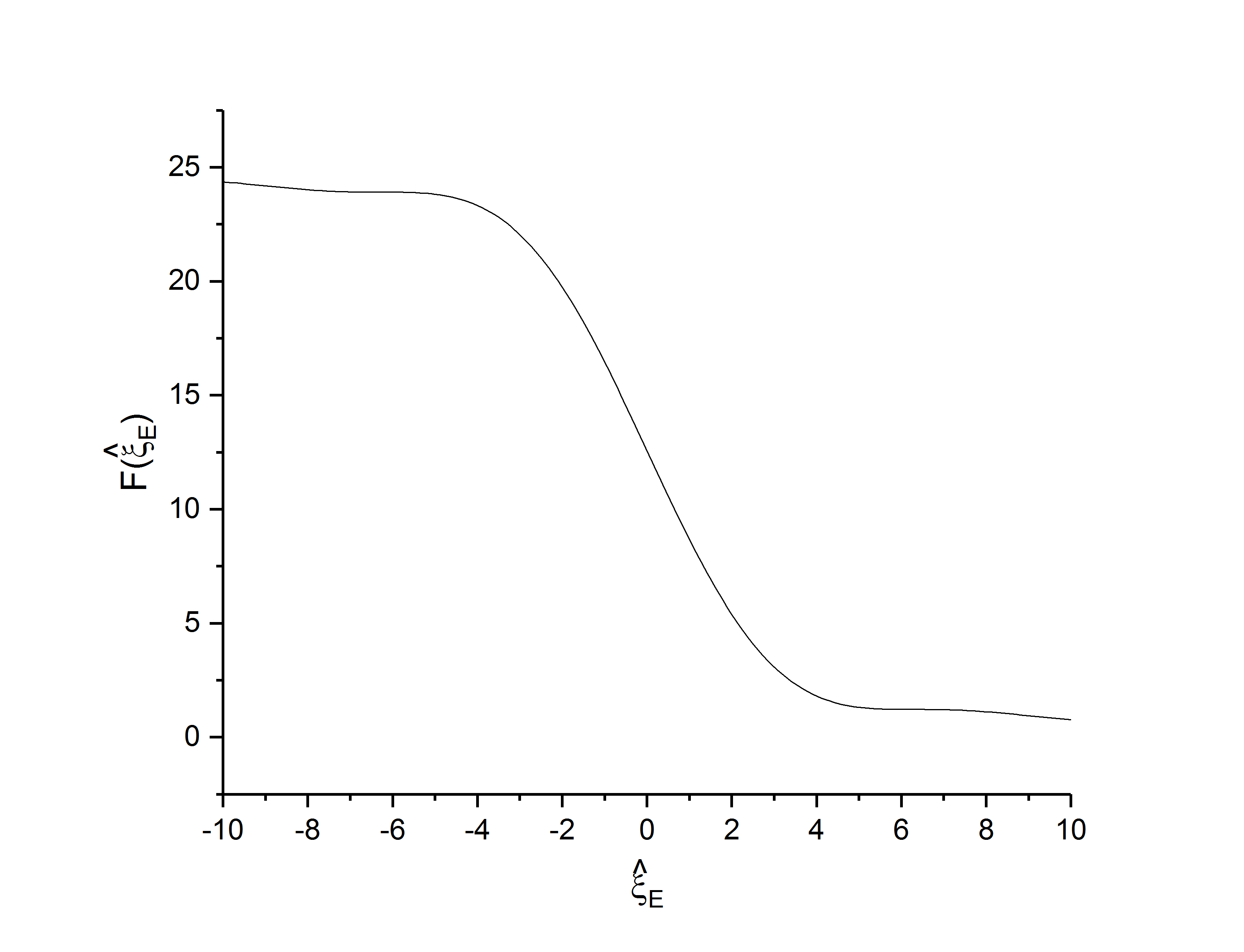}
\caption{Single electron. Left panel: calculated brightness as a function of $\hat{\xi}_E$ for parameters specified in the text. Right panel: the function $F(\hat{\xi}_E)$.}
	\label{BBBB}
\end{figure}

In order to make our argument clearer, we calculated the brightness for two specific cases using the code SPECTRA \cite{tanaka01}. Both cases refer to parameter compatible with the PETRA IV project, with an energy of $6$ GeV, and a planar undulator with period $\lambda_u = 65.6$ mm and a length of 5 m, corresponding to $76$ periods. We set zero emittance and discuss two single-electron cases with resonant energies at $580$ eV and $4000$ eV.   The results are plot in the left panel of Fig. \ref{BBBB} as a function of the detuning $\hat{\xi}_E=-4 \pi N_u ({\gamma-\gamma_1})/{\gamma_1}$, where we show the brightness  divided by the value at zero detuning. In the right panel of the same figure we plot the function $F(\hat{\xi}_E)$. By comparing the two plots one can see, as expected, a very similar behaviour. The only difference is that the brightness computed with SPECTRA (which is not based on the resonant approximation used for the analytical calculations) has its maximum around $\hat{\xi}_E \simeq -4$, while the analytical calculation shows that the function $F(\hat{\xi}_E)$ continues to grow for values of $\hat{\xi}_E$ below that. This last fact can be seen as a consequence of the fact that at the source, the maximum of the intensity profile is increasing for negative detuning values, see Fig. \ref{uno}, right panel, as previously discussed. Note that, in any case, the brightness is roughly anti-symmetrical with respect to the point $\hat{\xi}_E=0$ also for large detuning values, and this reinforces our conclusion that for zero emittance and symmetric energy spread distribution we cannot have any effect of the energy spread on the brightness, in agreement with the analysis of Eq. (\ref{Wigfarfinboh}) and Eq. (\ref{calcul}).

It is interesting to compare our results with those in \cite{tanaka09}. As discussed in the introduction, in \cite{tanaka09}   an approximated formula for the brightness is proposed, which is derived starting from Eq. (\ref{Busual}) that is the usual expression for the brightness based on Gaussian approximation, but includes the impact of a  modified spatial and angular profile of the undulator radiation in the presence of detuning.  In our notations, setting  for simplicity $N \equiv N_x = N_y$ and $D \equiv D_x = D_y$, this formula reads

\begin{eqnarray}
B_A =  B \left[  \frac{D}{\pi} + Q_a^2 \left(\frac{\sqrt{\Delta_E}}{2}\right)\right]^{-1}\left[4 \pi N + 4 Q_a^{4/3}\left(\frac{\sqrt{\Delta_E}}{8}\right)\right]^{-1}
\label{BA}
\end{eqnarray}
where

\begin{eqnarray}
Q_a(x) = \left[\frac{2 x^2}{-1 + \exp(-2x^2) + \sqrt{2\pi}~ x~ \mathrm{erf}(2^{1/2} x)}\right]^{1/2}
\label{xyz}
\end{eqnarray}
and the subscript "A" stands for "Approximated". 

\begin{figure}
	\centering
	\includegraphics[width=1.0\textwidth]{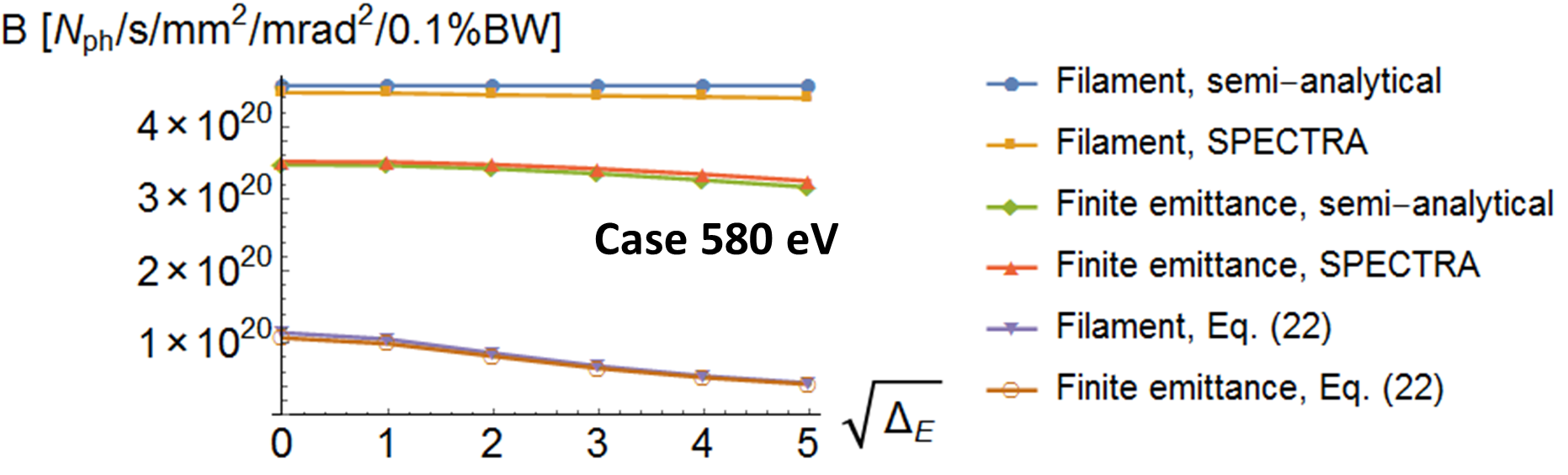}
	\includegraphics[width=1.0\textwidth]{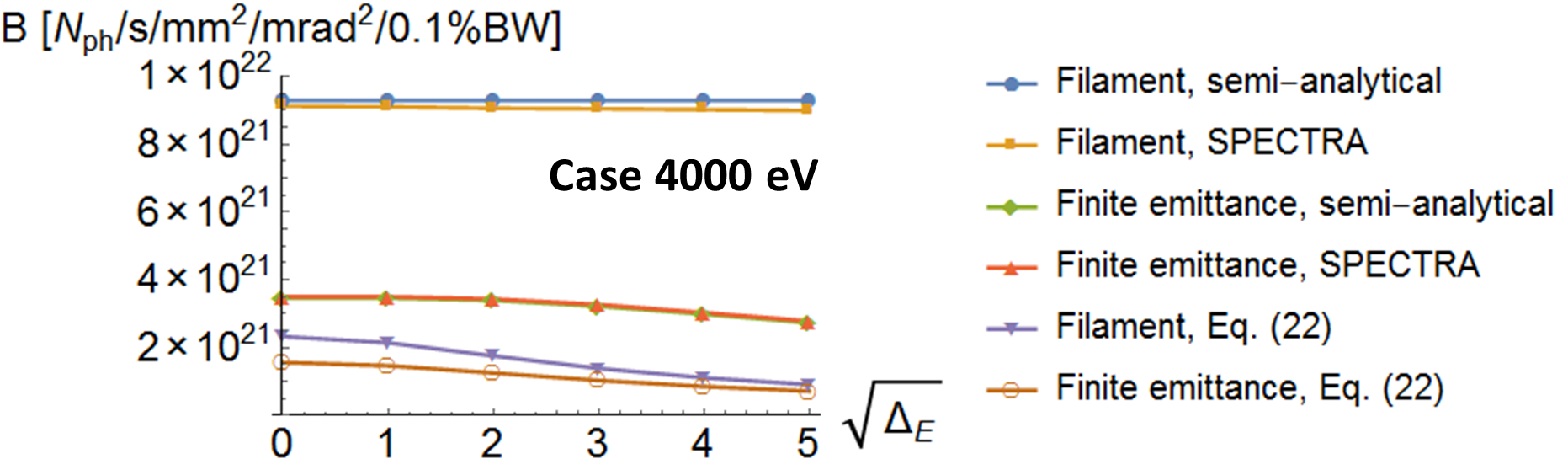}
	 \caption{A comparison of the brightness as a function of the energy spread for zero and non-zero emittance at two different resonant photon energies ($580$ eV and $4000$ eV, see text) and using different methods: Eq. (\ref{Wigfarfinboh}) (blue circles), Eq. (\ref{Wigfarfin00}) (green diamonds), SPECTRA calculations (orange squares and red upwards triangles) and  Eq. (\ref{BA}) (violet downwards triangles and brown empty circles).}
	\label{PcB}
\end{figure}
We considered the previously discussed parameters  compatible with the PETRA IV project and we analysed the case of zero emittance as well as the case for a finite emittance $\epsilon_{x,y}=10$ pm, equal betatron functions $\beta_{x,y}=1$ m, and no dispersion. A comparison of the brightness as a function of the energy spread for zero and non-zero emittance at the two different resonance photon energies of $580$ eV and $4000$ eV is shown in Fig. \ref{PcB}, as calculated using our formulas, SPECTRA, and Eq. (\ref{BA}). The cases of nonzero emittance correspond to ${N}_{x,y}=0.0059$ and ${D}_{x,y}=0.15$ for the case of $580$ eV, and ${N}_{x,y}=0.04$ and ${D}_{x,y}=1.01$ for the case of $4000$ eV. The main parameters are summarized in table \ref{table:sum}. Note that the detuning parameter depends linearly on the harmonic number. If we consider $N_u \simeq 100$ and $\sigma_{\Delta \gamma/\gamma} = 10^{-3}$, one immediately sees that the normalized rms energy spread parameter is about $1.3$ for the first harmonic, but since it scales linearly with the harmonic number, for the $5$th it would be already about $6.3$. Therefore, we chose to present plots up to values of $\sqrt{\Delta_E} = 5$.

\begin{table}
	\caption{Main parameters corresponding to the simulations in this paper}
	\begin{tabular}{c c c}
		\hline
		Parameter	&	Value	&	Unit	\\
		\hline
		$\epsilon_{x,y}$	&	10	&	pm	\\
		$\beta_{x,y}$	&	1	&	m	\\
		E	&	6	&	GeV	\\
		$\lambda_u$	&	65.6	&	mm	\\
		$N_u$ & 76 & - \\ 
		\hline
	\end{tabular}
	\label{table:sum}
\end{table}

Looking at Fig. \ref{PcB} we see that there is a factor $4$ difference between Eq. (\ref{BA})  in the limit for no emittance and energy spread and Eq. (\ref{Wigfarfinbohfin}). In \cite{tanaka09} this seems to be explained as due to the fact that while  the Gaussian approximation was used "to determine the angular divergence and source size, the spatial profile" was "derived by the spatial Fourier transform of the angular distribution of the complex amplitude", leading to a factor two in the source size. We argue that this procedure should not lead to any difference in the brightness in the case of zero emittance and energy spread, because in that limit one must have, \cite{kim87}:

\begin{eqnarray}
B = 4 F/\lambda^2
\end{eqnarray}
as is confirmed by Eq. (\ref{Wigfarfinbohfin}) and (see Fig. \ref{PcB}) by direct calculations with the code SPECTRA.  

Aside for the factor four discrepancy, we note that Eq. (\ref{BA}) approximates the influence of energy spread by summing emittance-related parameters ($N$ and $D$) with powers of the function $Q_a$ that depend on the energy-spread. Therefore, in the limit for zero emittance, energy spread effects dominate the brightness. In contrast to this,  Eq. (\ref{Wigfarfinboh}) is completely independent of the energy spread. This behaviour is exemplified in Fig. \ref{PcB}.

Our conclusion is that while Eq. (\ref{BA}) may constitute a good approximation in some region of the parameter space, when it comes to the limit for a diffraction-limited beam with non-negligible energy spread, a more detailed study is needed. In particular, when one is well within the diffraction limit, there is no region where the brightness is dominated by energy-spread effects.

Clearly, the above considerations are valid only for a vanishing emittance of the electron beam, i.e. in the limit for $D_{x,y} \ll 1$ and $N_{x,y} \ll 1$. In fact, even for vanishing offsets $N_{x,y} \ll 1$, if we cannot assume $D_{x,y} \ll 1$ the expression for the brightness includes the integrated spectral flux for electrons with different angles, and the sum of contributions with positive an negative detuning is now depending on the detuning value, at difference with the case above where $F(\hat{\xi}_E)+F(-\hat{\xi}_E) = 8 \pi$ , independently of $\hat{\xi}_E$.

\section{Effects of energy spread on coherence}

It is interesting to discuss possible effects of the energy spread on the coherence properties of undulator radiation. As before, we will first consider the case for zero emittance.

Clearly, the phase of the field in Eq. (\ref{fffiledn}) only depends on the electron offset, and is fully independent of $\hat{\xi}_E$, i.e. of $\gamma$. However, we note that the magnitude and, most importantly, the sign of the field depend on $\hat{\xi}_E$. Let us discuss the impact of this sign on the spectral degree of coherence. We write explicitly a simplified expression in the case of zero emittance as 

\begin{eqnarray}
&&g({\hat{\theta}}, \Delta {\hat{\theta}}) = \exp\left(-i \hat{z} {\hat{\theta}}{\Delta {\hat{\theta}}}\right) \mathcal{G}(\hat{\theta}, \Delta \hat{\theta})= \exp\left(-i \hat{z} {\hat{\theta}} {\Delta{\hat{\theta}}}\right) \cr && {\int_{-\infty}^{\infty} d \hat{\xi}_E ~\mathrm{sinc}\left(\frac{\hat{\xi}_E}{2} + \frac{(\hat{\theta}+\Delta \hat{\theta}/2)^2}{4}\right) \mathrm{sinc}\left(\frac{\hat{\xi}_E}{2} + \frac{(\hat{\theta}-\Delta \hat{\theta}/2)^2}{4}\right) \exp\left(-\frac{\hat{\xi}_E^2}{2\Delta_E}\right)}\cr &&\Bigg/\left\{\left[\int_{-\infty}^{\infty} d \hat{\xi}_E~ \mathrm{sinc}^2\left(\frac{\hat{\xi}_E}{2} + \frac{(\hat{\theta}+\Delta \hat{\theta}/2)^2}{4}\right)  \exp\left(-\frac{\hat{\xi}_E^2}{2\Delta_E}\right)\right]^{1/2}\right. \cr && \times \left.\left[\int_{-\infty}^{\infty} d \hat{\xi}_E~ \mathrm{sinc}^2\left(\frac{\hat{\xi}_E}{2} + \frac{(\hat{\theta}-\Delta \hat{\theta}/2)^2}{4}\right)  \exp\left(-\frac{\hat{\xi}_E^2}{2\Delta_E}\right)\right]^{1/2}\right\}\label{bigggg} \cr &&
\end{eqnarray}
This equation has been found on the basis of Eq. (\ref{big0}), where we took the limit for zero emittance and we assumed, for simplicity, that the two vectors $\bm{\hat{\theta}}$ and $\Delta \bm{\hat{\theta}}$ are directed along the same direction. This simplification does not deprive our model of any useful physics, but it makes all arguments scalars, and hence easier to consider. Further on, since, as remarked above, the phase of the field in Eq. (\ref{fffiledn}) only depends on the electron offset, we factorize $g$ in the product of $\mathcal{G}$ and of the phase factor $\exp(i \hat{\theta} \Delta \hat{\theta})$. Note that $\mathcal{G}$ is still allowed to assume negative values. 

It is easy to see by inspection of Eq. (\ref{bigggg}) that when $\sigma_E \rightarrow 0$, $G$ is different from unity, but $|g| = |\mathcal{G}| \longrightarrow 1$ everywhere. Moreover, on axis, i.e. for $\hat{\theta}=0$, one has $g= \mathcal{G} = 1$, while off-axis, i.e. for $\hat{\theta} \ne 0$, one has jumps of $\mathcal{G}$ from $+1$ to $-1$ at all those values of $\Delta \hat{\theta}$ where  $(\hat{\theta} + \Delta \hat{\theta}/2)^2/4$ and $(\hat{\theta} - \Delta \hat{\theta}/2)^2/4$ differ by an odd multiple of $\pi$. 

Let us consider the case of nonzero energy spread. If we look on-axis at  $\hat{\theta}=0$, from Eq. (\ref{bigggg}) we see directly that $g = \mathcal{G} = 1$. However, off-axis, an interesting phenomenon takes place. The field from different electrons with different detuning $\hat{\xi}_E$ experience a change in sign at different values of $\Delta \hat{\theta}$. This means that different electrons generate radiation with different wavefronts, and coherence is therefore decreased. This effect is encoded in the function $\mathcal{G}$, while the phase factor $\exp(i \hat{\theta} \Delta \hat{\theta})$ cannot change. To our understanding, this mechanism was not discussed before and is at the basis of any possible coherence deterioration related to energy spread effects. It is important to underline that it is present only off-axis, while energy spread alone cannot influence coherence properties on-axis. In the presence of a finite emittance,  one must include the effect of different angles $\bm{\hat{\eta}}$ in Eq. (\ref{big0}). Then, even on-axis, different electrons generate radiation with different wavefronts, and coherence deteriorates.

In order to illustrate our statements and to estimate the importance of the effects of energy spread on coherence we performed semi-analytical calculations for the case of zero emittance. We fixed different values of $\hat{\theta}$ and plot the cross-spectral density (a real function, in our case), the spectral degree of coherence, and the visibility calculated above in the far-zone as a function of $\Delta \hat{\theta}$ for different values of the energy spread.

\begin{figure}
	\centering
	\includegraphics[width=0.5\textwidth]{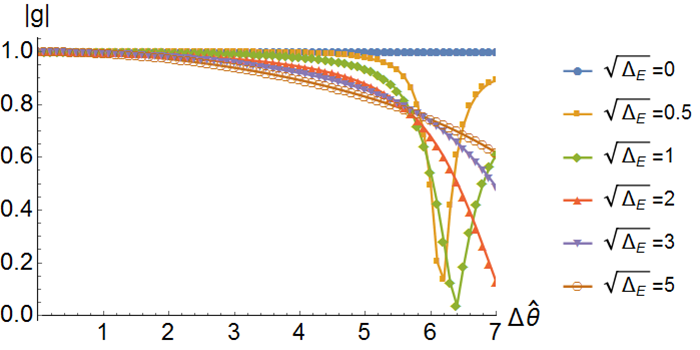}
	\includegraphics[width=0.5\textwidth]{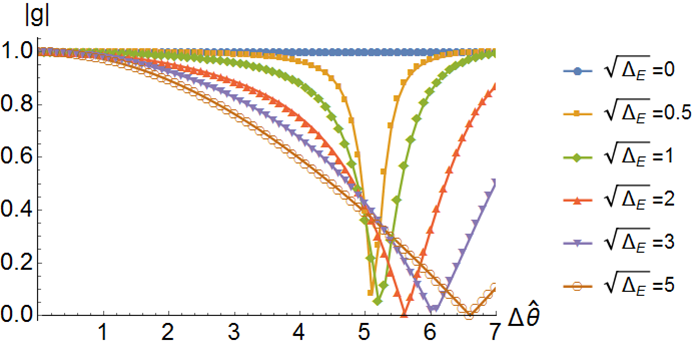}
	\\
	\includegraphics[width=0.5\textwidth]{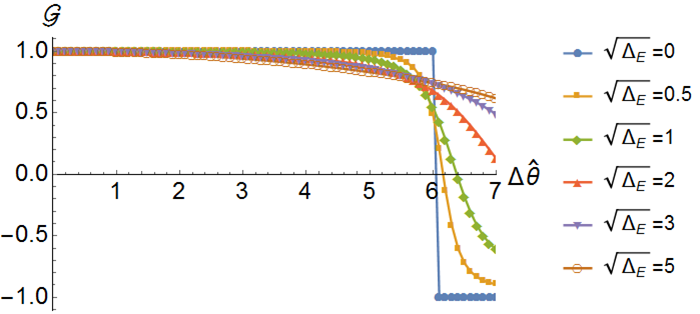}
	\includegraphics[width=0.5\textwidth]{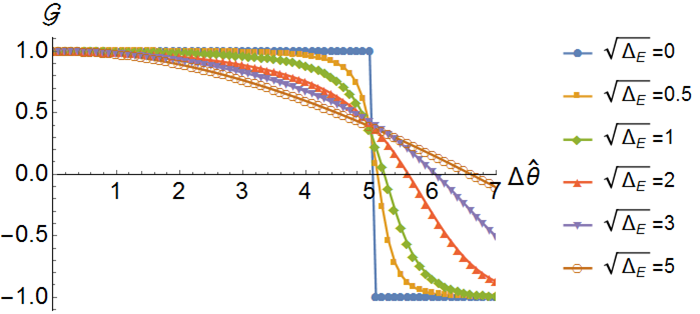}
	\\
	\includegraphics[width=0.5\textwidth]{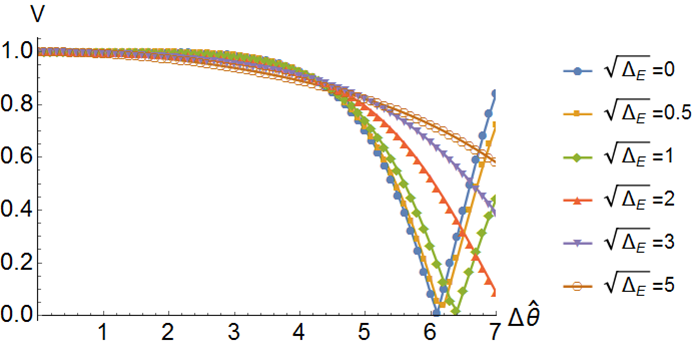}
	\includegraphics[width=0.5\textwidth]{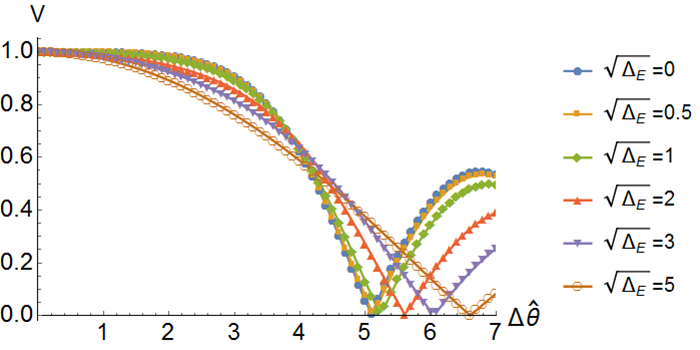}
	\\
	\includegraphics[width=0.6\textwidth]{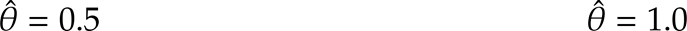}
	\caption{Far zone, zero emittance. Left panel: Modulus of the cross-spectral density, $|g|$ (top plot), the function $\mathcal{G}$ (middle plot) and the fringe visibility $V$ (lower plot) as a function of $\Delta \hat{\theta}$ for different values of the energy spread (see legend) and for $\hat{\theta} = 0.5$. Right panel: the same as in the left panel, for $\hat{\theta} = 1.0$. The symbols indicate actually simulated data. The solid lines are only a guide to the eye.}
	\label{due}
\end{figure}

Fig. \ref{due} presents results for $\hat{\theta}=0.5$ and $\hat{\theta}=1$. We remind the reader that the definition of our dimensionless units is given in Eq. (\ref{normq}). The normalization factor $\sqrt{\lambdabar/L_u}$ is of order of the angular size of the central cone. Therefore, it does not make too much sense to consider values of   $\hat{\theta}$ larger than unity.  As one immediately sees from the plots, even at $\hat{\theta}=1$ the effects of energy spread on coherence deterioration is very small. This is because the first change in sign for $\mathcal{G}$ happens at $\Delta \hat{\theta} = 2\pi$ (and the second would be at $\Delta \hat{\theta} = 4\pi$). Obviously there is little interest in going at such distance from the axis, and our conclusion is that the effect of energy spread on the deterioration of coherence is usually negligible in the far zone. 

However, the situation changes if the optics images, at the sample position, the virtual source in the middle of the undulator. In this case the previous analysis must be repeated using the quantities defined as before, but considering Eq. (\ref{nnfiledn}) instead of Eq. (\ref{fffiledn}).

\begin{figure}
	\centering
	\includegraphics[width=0.5\textwidth]{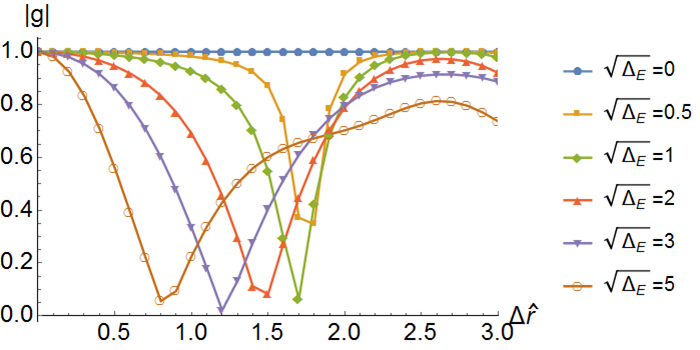}
	\includegraphics[width=0.5\textwidth]{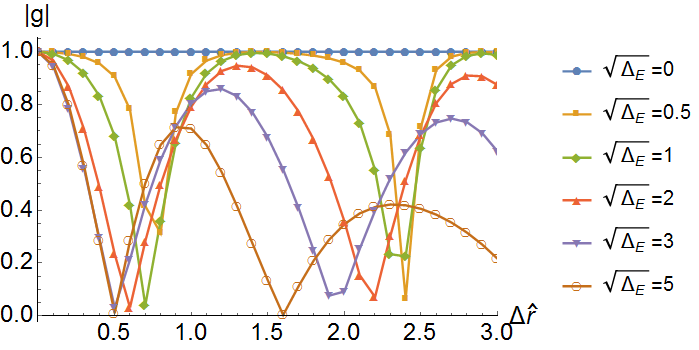}
	\\
	\includegraphics[width=0.5\textwidth]{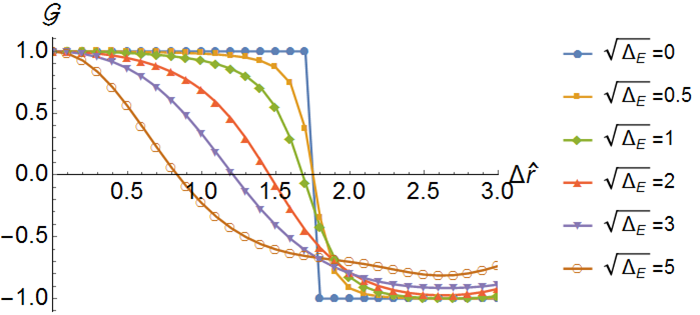}
	\includegraphics[width=0.5\textwidth]{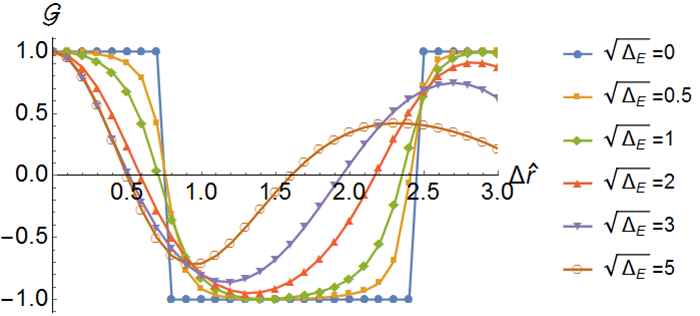}
	\\
	\includegraphics[width=0.5\textwidth]{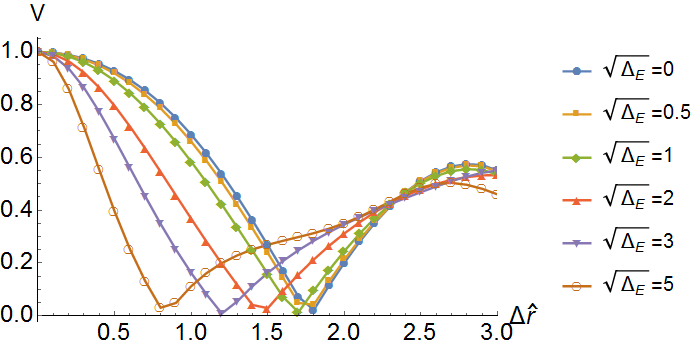}
	\includegraphics[width=0.5\textwidth]{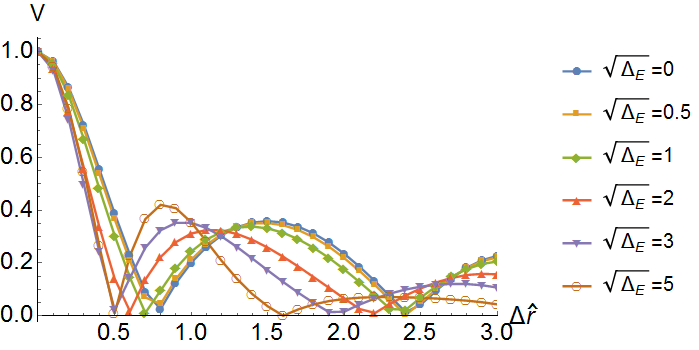}
	\\
	\includegraphics[width=0.6\textwidth]{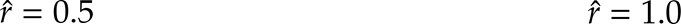}
	\caption{Virtual source, zero emittance. Left panel: Modulus of the cross-spectral density, $|g|$ (top plot), the function $\mathcal{G}$ (middle plot) and the fringe visibility $V$ (lower plot) as a function of $\Delta \hat{r}$ for different values of the energy spread (see legend) and for $\hat{r} = 0.5$. Right panel: the same as in the left panel, for $\hat{r} = 1.0$. The symbols indicate actually simulated data. The solid lines are only a guide to the eye.}
	\label{tre}
\end{figure}

Fig. \ref{tre} presents results for $\hat{r}=0.5$ and $\hat{r}=1$. This time, the normalization factor $\sqrt{\lambdabar L_u}$ in Eq. (\ref{normq}) is of the order of the transverse size of the central cone at the virtual source and, analogously as in the far zone, we limit ourselves to values of $\hat{r}$ up to unity. The same remarks made for the far zone hold for the values of the energy spread parameter. Inspection of Fig. \ref{tre} shows an important effect of the energy spread on coherence properties at the virtual source position. While we do not possess a simple expression as Eq. (\ref{fffiledn}) at the virtual source position, the mechanism that leads to coherence degradation is the same: namely, there is a change in the sign of the field, see Fig. \ref{uno}. This happens, however,  for smaller values of $\hat{r}$, which leads to degradation already for small values of $\Delta \hat{r}$, as seen from Fig. \ref{tre}.

It should be noted that, although the shape of the spectral degree of coherence is different when we compare the source with the far zone, the overall degree of coherence $\zeta$ remains unchanged. As discussed in the introduction, the overall degree of coherence is analogous to the trace of the square of the density matrix representing the statistical operator for a quantum mixture, the statistical operator being just, in our case, the cross-spectral density. The same degree of coherence can be expressed in terms of the Wigner distribution, because it is related to the cross spectral density by a simple Fourier transform:

\begin{eqnarray}
\zeta= \frac{\int d^2 \theta d^2 r ~W^2(\bm{r},\bm{\theta})}{\left[\int d^2 \theta d^2 r ~W(\bm{r},\bm{\theta})\right]^2}
\label{ovdeg}
\end{eqnarray}
Since the free-space propagation of the Wigner function is given by

\begin{eqnarray}
{W}(\bm{r},\bm{\theta};z)= {W}(\bm{r}-z \bm{\theta},\bm{\theta};0)
\label{Wprop}
\end{eqnarray}
a simple change of integration variables $\bm{r} \longrightarrow \bm{R}=\bm{r}-z \bm{\theta}$ shows that $\zeta$ is invariant for free-space propagation.\footnote{In principle one may directly show this fact in terms of integrals involving the spectral degree of coherence. However, carrying our the calculation explicitly would require extending the tabulation of Fig. \ref{due} and Fig. \ref{tre} to very large values of $\Delta \hat{\theta}$ and $\Delta \hat{r}$. We therefore prefer to give a synthetic, and more general demonstration of the invariance of $\zeta$.}

\begin{figure}
	\centering
	\includegraphics[width=0.5\textwidth]{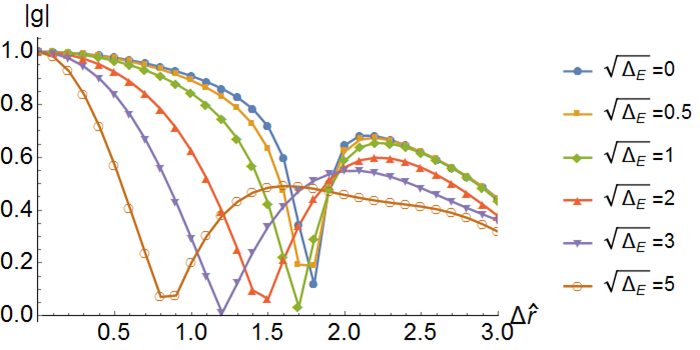}
	\includegraphics[width=0.5\textwidth]{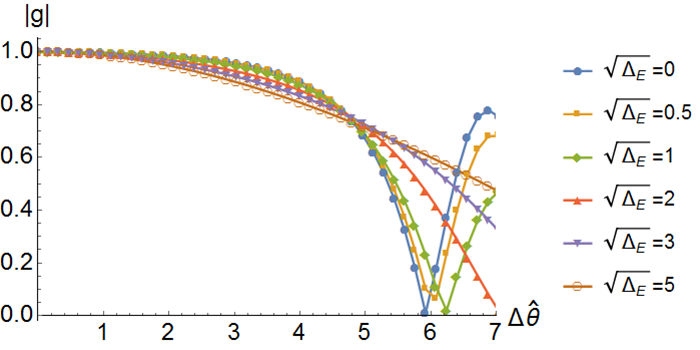}
	\\
	\includegraphics[width=0.5\textwidth]{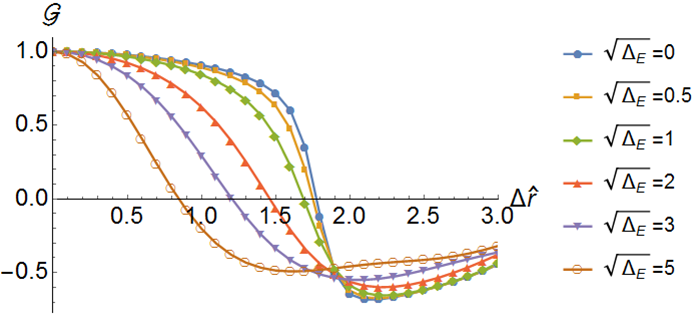}
	\includegraphics[width=0.5\textwidth]{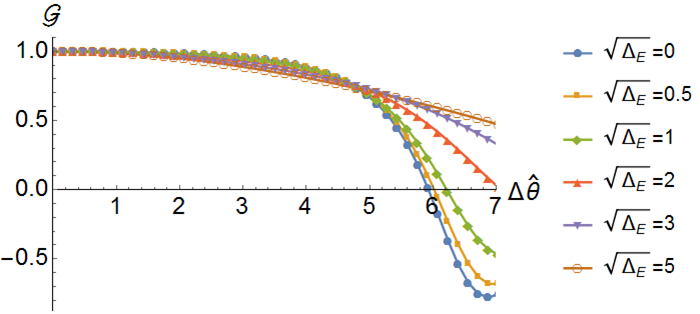}
	\\
	\includegraphics[width=0.5\textwidth]{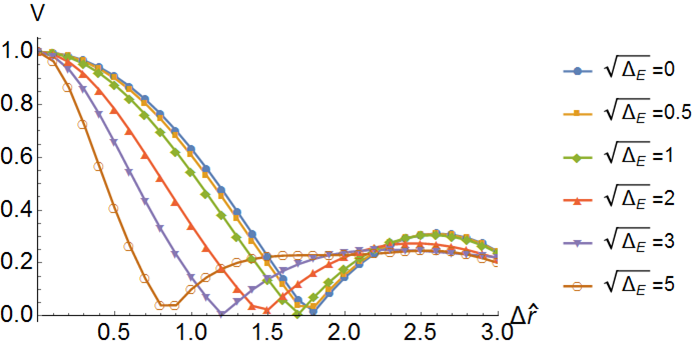}
	\includegraphics[width=0.5\textwidth]{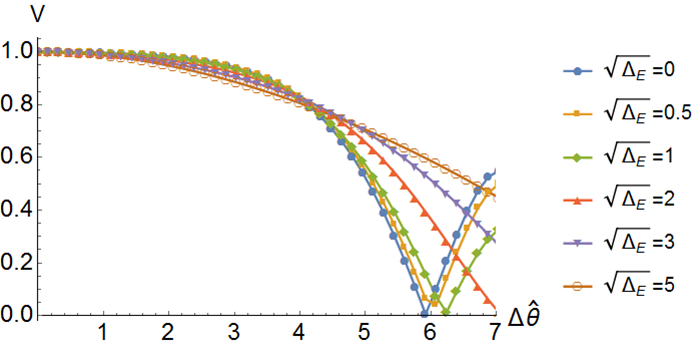}
	\\
	\includegraphics[width=0.6\textwidth]{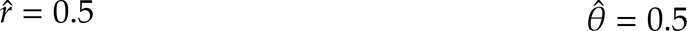}
	\caption{Non-zero emittance case with $N_x=N_y=0.0059$ and $D_x=D_y=0.15$ corresponding to the previously defined case for $580$ eV. Left side: Modulus of the cross-spectral density, $|g|$ (top plot), the function $\mathcal{G}$ (middle plot) and the fringe visibility $V$ (lower plot) as a function of $\Delta \hat{r}$ for different values of the energy spread (see legend) and for $\hat{r} = 0.5$ at the virtual source position. Right panel: the same as in the left panel, for $\hat{\theta} = 0.5$ in the far zone. The symbols indicate actually simulated data. The solid lines are only a guide to the eye.}
	\label{cinque}
\end{figure}

\begin{figure}
	\centering
	\includegraphics[width=0.5\textwidth]{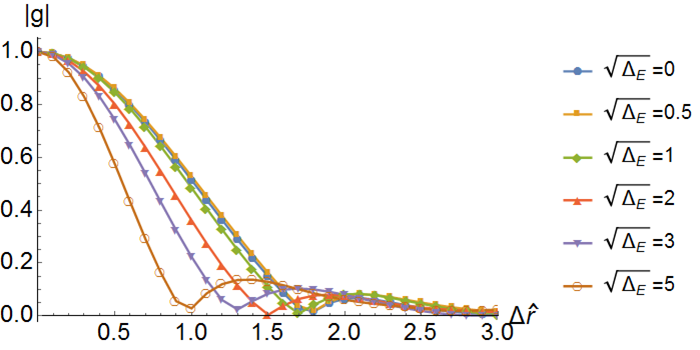}
	\includegraphics[width=0.5\textwidth]{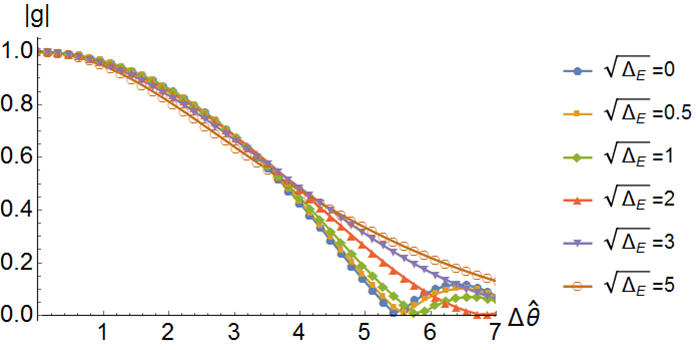}
	\\
	\includegraphics[width=0.5\textwidth]{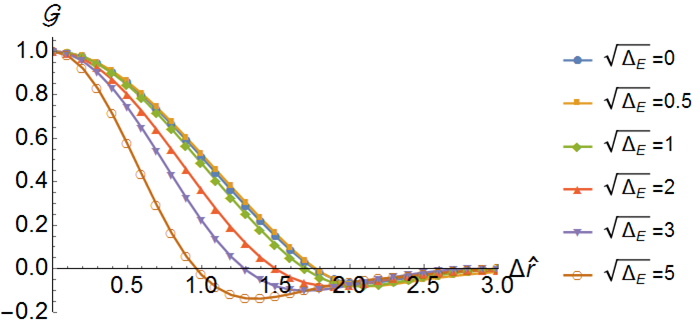}
	\includegraphics[width=0.5\textwidth]{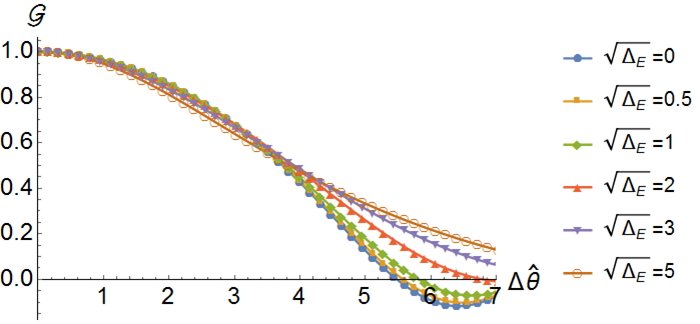}
	\\
	\includegraphics[width=0.5\textwidth]{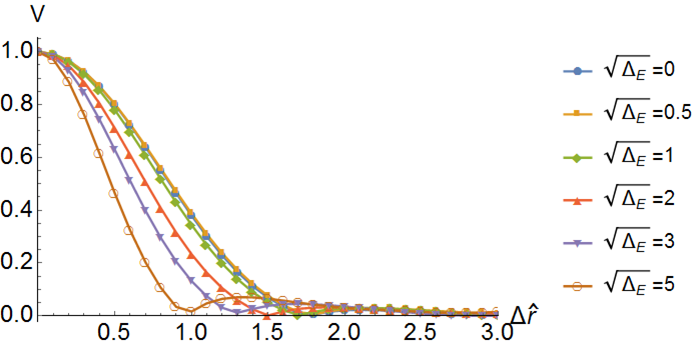}
	\includegraphics[width=0.5\textwidth]{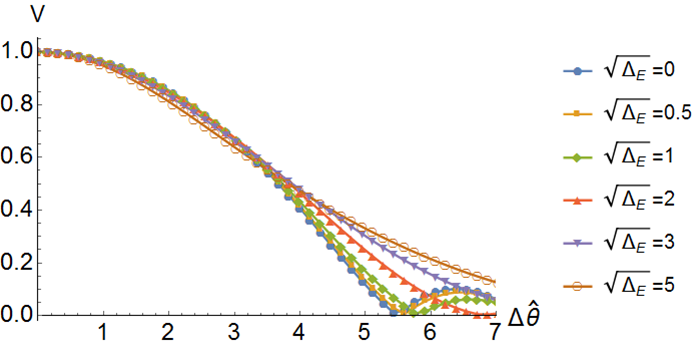}
	\\
	\includegraphics[width=0.6\textwidth]{Picture4_down.png}
	\caption{Non-zero emittance case with $N_x=N_y=0.04$ and $D_x=D_y=1.01$ corresponding to the previously defined case for $4000$ eV. Left side: Modulus of the cross-spectral density, $|g|$ (top plot), the function $\mathcal{G}$ (middle plot) and the fringe visibility $V$ (lower plot) as a function of $\Delta \hat{r}$ for different values of the energy spread (see legend) and for $\hat{r} = 0.5$ at the virtual source position. Right panel: the same as in the left panel, for $\hat{\theta} = 0.5$ in the far zone. The symbols indicate actually simulated data. The solid lines are only a guide to the eye.}
	\label{sei}
\end{figure}

%
We now complicate the situation by introducing finite emittance effects, corresponding to the two previously discussed cases respectively for $580$ eV and $4000$ eV. In particular we consider again the two settings $N_x=N_y=0.0059$, $D_x = D_y = 0.15$, corresponding to a resonant energy of $580$ eV, and $N_x=N_y=0.04$, $D_x = D_y = 1.01$, corresponding to a resonant energy of $4000$ eV. We set $\hat{\theta}=0.5$ in the far zone and $\hat{r} = 0.5$ at the virtual source position, and we plot the three functions $|g|$, $\mathcal{G}$ and $V$ at the virtual source and in the far zone. Results are shown in Fig. \ref{cinque} and Fig. \ref{sei}. Comparing Fig. \ref{tre} with Fig. \ref{cinque} and Fig. \ref{sei} we see how the effects of emittance become more and more important and finally dominate over energy spread effects. One can see coherence degradation already at zero energy spread, both at the virtual source and in the far zone. As is to be expected from the previous discussion,  energy spread effects are more visible at the virtual source, while in the far zone they are much less important.

It should be underlined once more that none of the degradation effects on coherence has an impact on the brightness when the beam has zero emittance. We checked this fact by using the expression for $\hat{G}$ to evaluate the brightness according to Eq. (\ref{Wigfarfin00}). No degradation was found in the case for zero emittance. However, as is obvious, in the case of non-zero emittance brightness degrades. Eq. (\ref{Wigfarfin00}) can, once more, be used to investigate the brightness degradation. 

The fact that the brightness cannot be affected by the energy spread alone, whereas the energy spread alone has an impact on the coherence properties of the radiation seems paradoxical. However, one should remember that in our case the brightness, according to our definition, is the Wigner distribution on-axis, i.e. at $\hat{r}=0$ and $\hat{\theta}=0$. As one can see from the previous analysis, at $\hat{r}=0$ and $\hat{\theta}=0$ there is no coherence degradation, whatever the energy spread parameter chosen. Intuitively speaking, the brightness is strictly related with the ability to focus a radiation beam on a sample. It can be spoiled by a decrease in spectral photon flux, by degradation of coherence or by wavefront distortions.  In the previous parts of this paper we showed that for a vanishing emittance and in the case of a symmetric energy spread distribution, one has a constant spectral photon flux over a large region of the energy spread parameter. However, we have seen here that there is an off-axis decrease of coherence. It is difficult to imagine that this has no effect on the ability to focus radiation. The dependence on the brightness on the on-axis Wigner function (where no coherence degradation takes place) seems, in this case, in contradiction with intuition. However, the decrease of off-axis coherence is only given by changes in sign of the field, happening at different transverse positions for particles with different energies. The ability to focus the field cannot depend of a change in sign, because it only introduces a trivial wavefront distortion: only trivial phase changes of $\pi$ are introduced by changing the energy, as in Fig. \ref{uno}. As a result, the brightness remains unvaried even though the coherence properties off-axis are degraded.

\section{Conclusions}

In this article we noted that changes in the brightness can be determined, roughly speaking, by influences related to the spectral photon flux, to the coherence, or to the wavefront. These three quantities can influence the brightness, because they influence the ability of focusing radiation onto the sample. Consider vanishing emittance and a symmetrical energy spread distribution. We have discussed a mechanism for degradation of coherence off-axis, while we have seen that, on-axis, coherence is preserved. Moreover, the field wavefront is not influenced (aside for a $\pi$ phase-difference) by the presence of energy spread, meaning that there cannot be any detrimental effect to the brightness, related with wavefront distortions. Finally, Eq. (\ref{calcul}) shows no effects on the flux, so we concluded that the brightness cannot be affected, in this case, by the energy spread. The same conclusion was reached by a direct calculation of the brightness, Eq. (\ref{Wigfarfinbohfin}). We studied the situation by means of semi-analytical calculations in Section \ref{sec:2}. 

In section \ref{sec:3} we extended our considerations to the case of a finite emittance. First, we confirmed our previous semi-analytical results. Then we increased the emittance and we studied its impact on coherence and brightness, showing how it degrades for parameters compatible with diffraction-limited storage rings of the next generation.
 
We conclude that there is no "energy-spread dominated" regime: when the emittance decreases, so does also the influence of the energy spread on coherence properties and brightness. 

The spectral degree of coherence is seen, instead, to decrease off-axis: this result is in agreement with our conclusion concerning the brightness. We illustrated our statements with simulation results, complementing them with remarks for the case of finite emittance.

\section{Acknowledgements}
We thank Ivan Vartaniants (DESY) for carefully reading the manuscript and for useful discussions on the subject.

\bibliographystyle{iucr}
\bibliography{iucr}

\end{document}